\documentclass[journal]{IEEEtran}
\usepackage[caption=false,font=footnotesize]{subfig}
\usepackage{bigints}
\usepackage{algorithm}
\usepackage{algpseudocode}
\usepackage{amsmath}
\usepackage{amsfonts}
\usepackage{amssymb}
\usepackage{amsthm}
\usepackage{mathtools}
\usepackage{color}
\usepackage[detect-all,detect-weight=true]{siunitx}
\usepackage{booktabs}
\usepackage{pgfplots}
\usepackage{cite}
\usepackage{textcomp}
\usepackage[hidelinks]{hyperref}
\usepackage[table]{colortbl}

\DeclareSIUnit\bit{b}

\usepgfplotslibrary{fillbetween, polar}
\usetikzlibrary{matrix, shapes.geometric, arrows, patterns} 
\pgfdeclarelayer{background}
\pgfsetlayers{background,main}

\tikzstyle{input} = [rectangle, rounded corners, minimum width=2cm, minimum height=1cm, text centered, draw=black, fill=orange!30]
\tikzstyle{output} = [rectangle, rounded corners, minimum width=2cm, minimum height=1cm, text centered, draw=black, fill=blue!30]
\tikzstyle{table} = [diamond, minimum width=2cm, minimum height=1cm, text centered, draw=black, fill=green!30]

\newcommand{\argmin}[1]{\underset{#1}{\operatorname{arg\,min\,}}}

\definecolor{lines-1}{RGB}{228,26,28}
\definecolor{lines-2}{RGB}{55,126,184}
\definecolor{lines-3}{RGB}{77,175,74}
\definecolor{lines-4}{RGB}{152,78,163}
\definecolor{lines-5}{RGB}{255,127,0}
\definecolor{lines-6}{RGB}{255,255,51}
\definecolor{lines-7}{RGB}{166,86,40}
\definecolor{lines-8}{RGB}{247,129,191}
\definecolor{lines-9}{RGB}{153,153,153}

\definecolor{verylightyellow}{RGB}{242,237,89}
\definecolor{verylightgray}{RGB}{200,200,200}
\definecolor{verylightred}{RGB}{255,191,191}
\definecolor{verylightgreen}{RGB}{117,234,123}

\pgfplotscreateplotcyclelist{myCycleList}{
	color=lines-1,thick,mark=o,mark size=3\\%
	color=lines-2,thick,mark=star,mark size=3\\%
	color=lines-3,thick,mark=square,mark size=3\\%
	color=lines-4,thick,mark=diamond,mark size=3\\%
	color=lines-5,thick,mark=triangle,mark size=3\\%
	color=lines-8,thick,mark=Mercedes star,mark size=3\\%
	color=lines-7,thick,mark=|,mark size=3\\%
	color=lines-9,thick,mark=oplus star,mark size=3\\%
}

\pgfplotsset{
	compat=1.14,
	width =\columnwidth, 
	height=.7\columnwidth,
	    ylabel absolute, ylabel style={yshift=-0.30cm},
	    xlabel absolute, xlabel style={yshift=0.2cm},
		label style={font=\footnotesize},
		tick label style={font=\footnotesize},
		legend style={font=\footnotesize},
		ymajorgrids=true,
		yminorgrids=true,
		minor grid style={dotted},
}

\makeatletter
\newcommand{\pushright}[1]{\ifmeasuring@#1\else\omit\hfill$\displaystyle#1$\fi\ignorespaces}
\newcommand{\pushleft}[1]{\ifmeasuring@#1\else\omit$\displaystyle#1$\hfill\fi\ignorespaces}
\newcommand{\subalign}[1]{%
	\vcenter{%
		\Let@ \restore@math@cr \default@tag
		\baselineskip\fontdimen10 \scriptfont\tw@
		\advance\baselineskip\fontdimen12 \scriptfont\tw@
		\lineskip\thr@@\fontdimen8 \scriptfont\thr@@
		\lineskiplimit\lineskip
		\ialign{\hfil$\m@th\scriptstyle##$&$\m@th\scriptstyle{}##$\crcr
			#1\crcr
		}%
	}
}
\newcommand{\vast}{\bBigg@{3.5}}
\makeatother

\begin{document}
	
\title{MmWave Hybrid Array with More Users than RF Chains}

\author{
	Nil~Garcia, Henk~Wymeersch,~\IEEEmembership{Member,~IEEE}, Christian~Fager,~\IEEEmembership{Senior Member,~IEEE}, Erik~G.~Larsson,~\IEEEmembership{Fellow,~IEEE},
	\thanks{
		N.~Garcia and H.~Wymeersch are with the Department of Signals and Systems, and C.~Fager is with the Department of Microtechnology and Nanoscience, Chalmers University of Technology, Gothenburg, Sweden.
		E.~G.~Larsson is with  the Division of Communication Systems, Department of Electrical Engineering (ISY), Link\"{o}ping University, Link\"{o}ping, Sweden.
		This research was supported in part, the VINNOVA COPPLAR project, funded under Strategic Vehicle Research and Innovation grant No.~2015-04849, 5Gcar, the Swedish Research Council (VR) and ELLIIT}
}

\maketitle

\begin{abstract}
	In millimeter wave communications, hybrid analog-digital arrays consisting of only a few radiofrequency (RF) chains offer an attractive alternative to costly digital arrays. A limitation to hybrid arrays is that the number of streams that can be transmitted simultaneously cannot exceed the number of RF chains. We demonstrate that if the states of the analog components and the symbols passed to the RF chains
	are jointly optimized on a symbol-by-symbol (SbS) basis, it is possible to achieve the same degrees of freedom as a digital array with linear precoding, thus effectively enabling the transmission of a large number of streams. To this end, an algorithm for SbS hybrid precoding is proposed based on a variation of orthogonal matching pursuit, particularly suited to the structure of our problem, which makes it fast and precise.
\end{abstract}

\section{Introduction}

Fifth generation (5G) cellular networks will deliver information rates in the order of gigabits per second by combining multiple advancements in the field of communications \cite{rappaport2013millimeter,wang2015multi}. For instance, signals shall be transmitted in the millimeter wave (mmWave) bands where large portions of underutilized spectrum are available. The small wavelength at mmWave also enables the compactification of large-scale arrays consisting of hundreds, or even thousands, of antenna elements into relatively small spaces \cite{roh2014millimeter}.
In fact large-scale multiple-input multiple-output (MIMO) systems are expected to play a pivotal role in 5G as they offer several communication improvements such as channel hardening, large array gains and better spectral efficiency \cite{rusek2013scaling,larsson2014massive}.

A major impediment for the implementation of large-scale MIMO at mmWave is the need to replicate many radio frequency chains (RFC), which increases the chip area, the power consumption and the overall cost \cite{swindlehurst2014millimeter,boccardi2014five,sun2014mimo}. To bypass these technological barriers, hybrid digital-analog arrays have been advocated as a possible remedy to fully digital architectures \cite{alkhateeb2014channel,han2015large}. The main idea is to shift some of the digital baseband processing towards the analog domain. Some examples of hybrid architectures are networks of phase shifters or switches \cite{mendez2016hybrid}, RF lenses \cite{brady2013beamspace}, arrays with low-resolution analog-to-digital and digital-to-analog converters (ADC/DAC) \cite{abbas2017millimeter}, and plasmonic arrays \cite{bonjour2016plasmonic}.

In hybrid arrays with a network of phase shifters, analog precoders are generated by changing the phases of the RF signals after conversion to intermediate or carrier frequency. By tweaking the phases appropriately, the streams of data can be transmitted towards arbitrary directions in space. A limitation of this type of architecture is that the number of streams that can be transmitted simultaneously is limited to the number of RFCs \cite{alkhateeb2014mimo,gao2015mmwave,heath2016overview}.
To break the limit on the maximum number of transmitted streams in the case of a phase shifting network, some architectural changes are required. In \cite{bonjour2016ultra,bonjour2018time}, the base station (BS) time-multiplexes multiple streams of data and transmits them through a single RFC. Then, it performs symbol-by-symbol beam steering in order to direct the beam of each symbol into a different direction. 
Their signal processing contribution is combined with a new hardware design based on plasmonics which allows ultra fast phase shifting.
Another interesting approach is that of \cite{he2018spatial} where the number of transmitted streams is effectively increased by one by turning on and off some of the antennas.

This work proposes a hybrid precoding technique for transmitting more streams than the number of RFCs. Contrary to standard hybrid precoding, the analog and digital precoders are both changed on a symbol-by-symbol (SbS) basis. Specifically, we show that SbS precoding can imitate the transmitted signals of a digital array with linear precoding, effectively allowing the BS to transmit simultaneously a large number of streams. To enable SbS precoding we adopt the hybrid architecture of \cite{yu2018hardware} which uses switches connected to fixed phase shifters because they are cheaper and easier to implement in hardware.

The proposed algorithm for computing the SbS precoder is based on orthogonal matching pursuit (OMP) which is a well established tool from compressive sensing. However, contrary to previous works which had to rely on suboptimal dictionaries \cite{venkateswaran2010analog,el2012low,alkhateeb2013hybrid,de2016hybrid}, we show that it is possible to use a dictionary that includes the set of all possible phase shifts without increasing the computational complexity of OMP. The main drawbacks of the proposed strategy are (i) tight synchronization is required between the transmitted symbols and the switches, and (ii) spectral regrowth and inter-symbol interference can occur without adequate pulse shaping.

\section{System Model} \label{sec_signal_model}

\subsection{Architecture}

\begin{figure}
	\centering
	\includegraphics[width=\columnwidth]{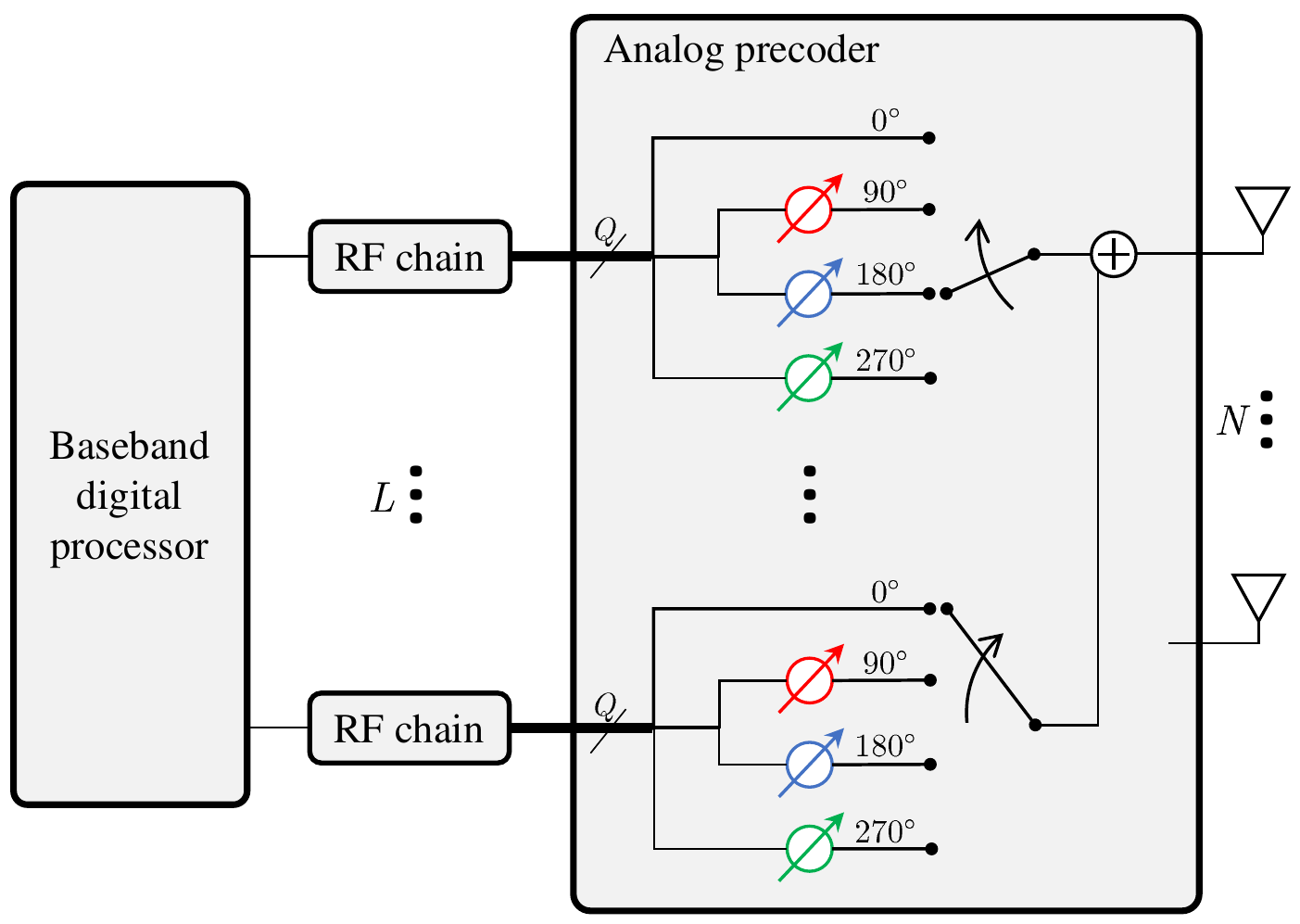}
	\caption{Uniform Linear Array (ULA) with $N$ antennas and $L$ radio-frequency chains (RFC). The phase shifters in the same color form an $L$-channel phase shifter (PS) that can simultaneously process $L$ signals in parallel. There are $Q-1$ multi-channel PS, where $Q$ is the number of different phase shifts (including the \SI{0}{\degree}-shift). The connection between each RFC and antennas includes a switch of $Q$ positions that selects the appropriate phase.}
	\label{fig:architecture}
\end{figure}
This work assumes the hybrid array architecture proposed in \cite{yu2018hardware} and portrayed in Fig.~\ref{fig:architecture} consisting of $N$ antennas, $L\ll N$ RFCs, and a network combining phase shifters and switches. Each RFC's output is connected to $Q$ fixed phase shifters ($Q-1$ if the \SI{0}{\degree}-phase shifter is included) with $Q$ different equispaced phases. In general, $Q$ is much smaller than the number of antennas (e.g., $Q=4$ in Fig.~\ref{fig:architecture}) and dictates the resolution of the phase shifters (\SI{90}{\degree} in Fig.~\ref{fig:architecture}). Each fixed phase shifter is in fact an $L$-channel phase shifter (marked with same color in Fig.~\ref{fig:architecture}) that processes the outputs of all RFCs in parallel. A dynamic switching network is cascaded after the phase shifters allowing each antenna to combine the outputs of each RFC with different phase shifts. 
While offering the same degree of flexibility as a hybrid array with a fully connected networks of phase shifters \cite{el2013multimode,han2015large,mendez2016hybrid,park2017dynamic}, its hardware complexity is substantially lower because it employs only $Q$ multi-channel fixed phase shifters, and adaptive switches are much easier to implement than adaptive phase shifters.

\subsection{Hybrid Precoding}

We assume a single-carrier narrow-band array, where the transmitted signal's bandwidth is much smaller than the carrier frequency \cite{cudak2014experimental}.
Let
\begin{equation} \label{eq:set_of_phases}
	\mathcal{P}=\{1,e^{i 2\pi /Q}\ldots,e^{i 2\pi (Q-1) /Q}\}
\end{equation}
be the set of fixed discrete phase shifts, and denote $\mathbf{s}_t=[s_{1,t}\cdots s_{L,t}]^\mathrm{T}$ the complex-valued symbols of the $L$ streams passed to the RFCs at time $t$.
Then, the transmitted baseband signal across the $N$ antennas at discrete time $t$ is
\begin{equation} \label{eq:signal_hybrid}
	\mathbf{y}_t =
	\mathbf{F}^{\text{RF}} \mathbf{s}_t
\end{equation}
where $\mathbf{F}^{\text{RF}}\in\mathcal{P}^{N\times L}$ is termed the analog precoding matrix,
\begin{equation}
	\mathbf{F}^{\text{RF}} \triangleq
	\begin{bmatrix}
		\mathbf{f}_1^{\text{RF}} s_{1,t} & \cdots & \mathbf{f}_L^{\text{RF}} s_{L,t}
	\end{bmatrix},
\end{equation}
and $\mathbf{f}_l^{\text{RF}}\in\mathcal{P}^{N\times 1}$ for $l=1,\ldots,L$, are the analog precoders associated to streams $1,\ldots,L$, respectively.
Standard hybrid precoding limits the number of transmitted streams to $L$. In the next section, we propose a symbol-by-symbol (SbS) precoding technique that enables the transmission of $K$ streams such that $L<K\leq N$.

\subsection{Problem Formulation} \label{sec:problem_formulation}

We aim to develop an algorithm to support any number of streams only limited by the number of antennas, like in the case of a digital array. To this end, we allow the switches to change their states on a symbol-by-symbol basis, effectively making the analog precoders in \eqref{eq:signal_hybrid} a function of time
\begin{equation} \label{eq:signal_hybrid:SbS}
	\mathbf{y}_t^\text{hyb} =
	\mathbf{F}_t^{\text{RF}} \mathbf{f}_t^\text{BB},
\end{equation}
where $\mathbf{f}_t^\text{BB}$ plays the role of $\mathbf{s}_t$, but different notation is used because as it will be apparent later, these symbols do not represent data.
In the case of a digital array transmitting $K$ streams, the baseband signal across all antennas is
\begin{equation} \label{eq:signal_digital}
	\mathbf{y}_t^\text{dig} = \sum_{k=1}^{K}\mathbf{f}_k^\text{dig} s_{k,t} = \mathbf{F} \mathbf{s}_t,
\end{equation}
where $\mathbf{f}_k^\text{dig}\in\mathbb{C}^{N\times 1}$ is the digital precoder for stream $k$, $\mathbf{F}=\begin{bmatrix}\mathbf{f}_1 & \cdots & \mathbf{f}_K\end{bmatrix}$ and $\mathbf{s}_t=\begin{bmatrix}s_{1,t} & \cdots & s_{K,t}\end{bmatrix}^\mathrm{T}$.
The goal is to find a method that makes
\begin{equation} \label{eq:approximation}
	\mathbf{F}_t^\text{RF} \mathbf{f}_t^\text{BB} \approx \mathbf{y}_t^\text{dig},
\end{equation}
thus, making the hybrid array effectively transmit the same signal than a digital array.

\paragraph*{Remark} The effect of ultra fast switching on the signal has not been taken into consideration in this work. The signals going through the phase shifters and switches are commonly shaped by square-root raised cosine filters, which span multiple symbols in time. Fast switching between symbols will create sharp transitions on the signals, causing inter-symbol interference and spectral regrowth. Techniques for solving or mitigating this problem are left for future study. Possible solutions may be to include passband filters at the antennas or employ pulses of duration less than a symbol as in \cite{bonjour2018time}.

\section{Proposed Approach} \label{sec:more_streams}

The main idea of the proposed symbol-by-symbol (SbS) precoding scheme is to change the analog and digital precoders at the symbol rate such that the baseband signals transmitted by the hybrid and digital arrays are \emph{the same} \eqref{eq:approximation}.
A simple approach is to solve the least squares fit
\begin{equation} \label{opt:hybrid_precoding:LS}
	\min_{\subalign{\mathbf{F}_t^\text{RF}&\in\mathcal{P}^{N\times L}\\\mathbf{f}_t^\text{BB}&\in\mathbb{C}^{L\times1}}}
	\left\|
		\mathbf{y}_t^\text{dig} - \mathbf{F}_t^\text{RF} \mathbf{f}_t^\text{BB}
	\right\|_2,
\end{equation}
for all $t$, where $\mathbf{y}_t^\text{dig}$ is defined in \eqref{eq:signal_digital}.
If the obtained solution is zero, then the SbS precoder perfectly approximates the digital precoder at time $t$.
Nonetheless, the are several challenges to solving \eqref{opt:hybrid_precoding:LS}.
\begin{itemize}
	\item[A)] The optimization is over discrete variables, $\mathbf{F}_t^\text{RF}$, for which, in general, the global optimum cannot be retrieved efficiently.
	\item[B)] The solver must enjoy low computational complexity since this problem needs to be solved for each symbol time $t$, even when the number of antennas, $N$, grows very large.
	\item[C)] For a good approximation \eqref{eq:approximation}, the optimum value must be close to zero for the entire duration of the stream.
\end{itemize}
The next section proposes some algorithms for accurate and fast SbS precoding and their performance is evaluated numerically in Section~\ref{sec:numerical_results}.

\subsection{Sparse Reformulation}

To avoid overloading the notation, the subindex $t$ is dropped from hereon.
Solving the mixed discrete-continuous optimization problem \eqref{opt:hybrid_precoding:LS} is challenging. An alternative approach \cite{alkhateeb2014channel} is to recast it as a sparse recovery problem
\begin{equation} \label{opt:CS}
\begin{split} 
\min_{\mathbf{x}}
&\quad\left\lVert\mathbf{y}^\text{dig} - \mathbf{\Omega} \mathbf{x} \right\rVert_2\\
\text{s.t.}
&\quad\left\lVert\mathbf{x} \right\rVert_0 \leq L,
\end{split}
\end{equation}
where $\lVert\cdot\rVert_0$ counts the number of non-zero entries, $\mathbf{x}$ is an $L$-sparse vector, and the dictionary matrix $\mathbf{\Omega}\in\mathcal{P}^{N\times M}$ is built by concatenating in columns all vectors in $\mathcal{P}^{N\times 1}$, i.e., 
\begin{equation} \label{eq:dictionary:complete}
\mathbf{\Omega} =
\begin{bmatrix}
\boldsymbol{\omega}_1 & \cdots & \boldsymbol{\omega}_M
\end{bmatrix},
\end{equation}
where $\left\{\boldsymbol{\omega}_1, \ldots, \boldsymbol{\omega}_M\right\}=\mathcal{P}^{N\times 1}$.
Thus, the number of columns in the dictionary is $M=|\mathcal{P}^{N\times 1}|=Q^{N}$ where $Q=|\mathcal{P}|$ is the number of different phase shifts. 
Assuming $\mathbf{x}$ is the global optimum, $\mathbf{f}^\text{BB}$ is obtained by keeping only the non-zero entries of $\mathbf{x}$, and $\mathbf{F}^\text{RF}$ is formed by selecting the columns of $\mathbf{\Omega}$ with the same indexes as the non-zeros of $\mathbf{x}$.

In the above sparse setting, vector $\mathbf{x}$ can then be recovered using compressive sensing tools. Because sparse problems are NP-hard, retrieval of the global optimum is not guaranteed, but they can still lead to fairly good solutions. 
Orthogonal matching pursuit (OMP) offers a good trade-off between computational complexity and quality of the solution. However, because in our case the dictionary size grows exponentially with the number of antennas, normal OMP is not a feasible approach. In the next section we delve into these issues and propose an efficient implementation of OMP.

\subsection{The Naive OMP} \label{sec:naive}

OMP is a greedy method which attempts to solve \eqref{opt:CS} by finding the $L$ non-zero entries of $\mathbf{x}$ recursively.
At iteration $l\in\{1,\ldots,L\}$ OMP solves the following simplified problem
\begin{equation} \label{eq:CS_simple}
\min_{\substack{\beta\in\mathbb{C}\\\boldsymbol{\omega}\in\mathcal{P}^{N\times 1}}}
\left\lVert\mathbf{r}_{l-1} - \beta \boldsymbol{\omega} \right\rVert_2,
\end{equation}
where $\mathbf{r}_{l-1}$ is a residual vector from the previous iteration (initialized to $\mathbf{r}_{0}=\mathbf{y}^\text{dig}$). By plugging back the least squares estimate of the nuisance parameter $\hat{\beta}=\boldsymbol{\omega}^\mathrm{H} \mathbf{r}_{l-1} \|\boldsymbol{\omega}\|_2^{-2}$, and using the fact that the norm of $\boldsymbol{\omega}$ is constant because all elements have unit-magnitude, \eqref{eq:CS_simple} transforms to
\begin{equation} \label{eq:OMP:max_corr}
\boldsymbol{\omega}^\star = \argmin{\boldsymbol{\omega}\in\mathcal{P}^{N\times 1}}
\left|
\boldsymbol{\omega}^\mathrm{H} \mathbf{r}_{l-1}
\right|.
\end{equation}
The newly estimated vector of phase shifts $\boldsymbol{\omega}^\star$ is concatenated with the previous ones through $\mathbf{B}_l= [\mathbf{B}_{l-1} \, \boldsymbol{\omega}^\star]\in\mathcal{P}^{N\times l}$, and a new residue is obtained by orthogonalizing the target precoder, i.e.,
\begin{equation} \label{eq:residue}
\mathbf{r}_l = \operatorname{Proj}_{\mathbf{B}_l}^\perp \mathbf{y}^\text{dig},
\end{equation}
where $\operatorname{Proj}_{\mathbf{B}_l}^\perp = \mathbf{I} -\mathbf{B}_l(\mathbf{B}_l^\mathrm{H}\mathbf{B}_l)^{-1} \mathbf{B}_l^\mathrm{H}$. Upon completion of the $L$-iteration, we assign $\mathbf{F}^\text{RF}=\mathbf{B}_L$ and $\mathbf{f}^\text{BB}=(\mathbf{B}_L^\mathrm{H}\mathbf{B}_L)^{-1} \mathbf{B}_L^\mathrm{H}\mathbf{y}^\text{dig}$.
For every iteration, \eqref{eq:OMP:max_corr} is solved by computing the matrix-vector product $\mathbf{\Omega}^\mathrm{H}\mathbf{r}_{l-1}$ with computational complexity\footnote{Check \cite{cormen2009introduction} for the computational complexity of different operations.} $\mathcal{O}(NM)$.
The bottleneck in computing the residue \eqref{eq:residue} is caused by the matrix product $\mathbf{B}_l^\mathrm{H}\mathbf{B}_l$, whose computational complexity is $\mathcal{O}(Nl^2)$. Thus, the total complexity of running iteration $l$ is $\mathcal{O}(NM +Nl^2)$.

In the case of a dictionary including all possible phase shifts, the number of columns becomes $Q^{N}$, making the implementation of naive OMP inviable due to its exponential complexity $\mathcal{O}(N Q^{N} +Nl^2)$. Next, we present the \emph{steering dictionary}. A simple fix in the mmWave literature on hybrid precoding \cite{venkateswaran2010analog,el2012low,alkhateeb2013hybrid,alkhateeb2014channel} consists in reducing the dictionary to just a few columns.
In particular, in \cite{de2016hybrid,alkhateeb2013hybrid,el2012low} the target precoder $\mathbf{y}^\text{dig}$ is assumed to be a linear combination of $L$ or less steering vectors:
$\mathbf{y}^\text{dig}=\sum_{k=1}^{K\leq L} w_k\mathbf{a}(\theta_k)$. Therefore, solving \eqref{opt:CS} only requires a dictionary that contains steering vectors,
\begin{equation} \label{eq:dictionary:steer}
\mathbf{\Omega} = 
\begin{bmatrix}
1 & 1 & \cdots & 1 \\
1 & e^{i\frac{2\pi}{Q}} & \cdots & e^{i(Q-1)\frac{2\pi}{Q}} \\
\vdots & \vdots & \cdots & \vdots \\
1 & e^{i(N-1)\frac{2\pi}{Q}} & \cdots & e^{i(N-1)(Q-1)\frac{2\pi}{Q}}
\end{bmatrix},
\end{equation}
greatly reducing the dictionary size. Every column in \eqref{eq:dictionary:steer} belongs to the array manifold and all entries in the dictionary are discrete phase shifts in $\mathcal{P}$.
Operating with such dictionary has the advantage that the computational complexity of naive OMP is now polynomial $\mathcal{O}(NQ +Nl^2)$ because $M=Q$.

\subsection{The Complete Dictionary}

The main shortcoming of the steering dictionary (or other dictionaries that only include a subset of all possible phase shifting vectors) in the previous section is that, in general, it works only if the target precoder can be represented as a linear combination of $L$, or less, steering vectors. Obviously, this is not the case when the number of streams exceeds the number of RFCs, i.e., $K>L$.
Next, we show how to run OMP efficiently using the complete dictionary containing all possible phase shifts. 

In OMP, the dictionary only plays a role when finding the vector of phase shifts that correlates the most with the residue \eqref{eq:OMP:max_corr}. Computing the scalar product between the residue and each column with the complete dictionary is completely inviable for large arrays because its size grows exponentially with the number of antennas.
Nevertheless, it turns out that it can be solved without explicitly computing these scalar products upon noticing that it is equivalent to the problem of multiple-symbol differential detection in communications, for which the efficient Algorithm~\ref{alg:optimal_phases} was proposed in \cite{mackenthun1994fast}, and later rediscovered in \cite{sweldens2001fast}. A description is provided next without delving in the proofs.
\begin{algorithm} \caption{Optimal phase shifts} \label{alg:optimal_phases}
	\begin{algorithmic}[1]
		
		\Statex \textbf{inputs:} $\mathbf{r}$ and $\delta$
		
		\Statex \textbf{outputs:} $\boldsymbol{\omega}$
		
		\State set $z_n = |r_n| e^{i \operatorname{mod}(\angle r_{n},\delta)}$ for $n=1,\ldots,N$ \label{alg:optimal_phases:init}
		
		\State let $\Omega$ be an ordering such that $\angle z_{\Omega(1)}\leq \ldots \leq \angle z_{\Omega(N)}$ \label{alg:optimal_phases:sort}
		
		\State initialize $S=\sum_{n=1}^{N}z_{n}$, $v=|S|$ and $m_\text{opt}=0$  \label{alg:optimal_phases:init_sum}
		
		\For{$m=1,\ldots,N-1$} \label{alg:optimal_phases:for}
		
		\State $S \leftarrow S -z_{\Omega(m)} + z_{\Omega(m)}e^{i\delta}$ \label{alg:optimal_phases:end_loop}
		
		\If{$|S|>v$} \label{alg:optimal_phases:begin_loop}
		
		\State $v=|S|$ and $m_\text{opt}=m$
		
		\EndIf
		
		\EndFor \label{alg:optimal_phases:end_for}
		
		\State $\omega_{\Omega(n)} = \begin{cases} e^{i\left(\angle r_{\Omega(n)} -\angle z_{\Omega(n)} -\delta\right)} & n=1,\ldots,m_\text{opt} \\ 
		e^{i\left(\angle r_{\Omega(n)} -\angle z_{\Omega(n)}\right)} & n=m_\text{opt}+1,\ldots,N \end{cases}$ \label{alg:optimal_phases:output}
		
	\end{algorithmic}
\end{algorithm}
Let $\angle x$ denote the phase of $x$ in radians, and define the phase shifters' resolution as $\delta\triangleq\frac{2\pi}{Q}$.
For an intuitive explanation on the algorithm, first, note \eqref{eq:OMP:max_corr} can also be expressed as $\arg\max |\sum_n z_{n}|$,
where $z_n \triangleq \omega_n^* \cdot r_{l-1,n}$ for all $n$.  
Let $z^\star_n$ and $\omega^\star_n$ be the optimal $n$-th components. Then, for all $n$, the magnitudes of $z^\star_n$ and $r_{l-1,n}$ must match, and the phase difference between $z^\star_n$ and $r_{l-1}$ must be a multiple of $\delta$.
Assuming $\mathbf{z}^\star\triangleq[z_1^\star \cdots z_N^\star]$ is found, $\boldsymbol{\omega}^\star$ can easily be obtained by reversing the variable change since $\mathbf{r}_{l-1}$ is a known input parameter. It turns out that $\mathbf{z}^\star$ satisfies the property that the phase difference\footnote{This is the wrap-around phase difference, e.g., $|\angle 0- \angle 5 e^{j\frac{3\pi}{2}}|_\text{wa}=\frac{\pi}{2}$.} between any two components must be smaller than the phase shifters' angle resolution \cite{mackenthun1994fast,sweldens2001fast}, i.e.,
\begin{equation} \label{eq:phase_diff}
|\angle z_n^\star- \angle z_m^\star|_\text{wa} <\delta \text{ for any }n,m.
\end{equation}
Intuitively, by having similar phases, the entries of $\mathbf{z}^\star$ add up constructively and maximize $|\sum_n z_{n}|$.
An example of a point satisfying \eqref{eq:phase_diff} is given in line~\ref{alg:optimal_phases:init} of Algorithm~\ref{alg:optimal_phases}. This point is essentially a copy of $\mathbf{r}_{l-1}$ whose phases have been rotated multiples of $\delta$ until they all fall within a sector of width $\delta$. Moreover, it can be further shown, that there are at most $N$ different\footnote{Different in the sense of being pair-wise linearly independent.} points, say $\{\mathbf{z}^{(1)},\ldots,\mathbf{z}^{(N)}\}$, satisfying \eqref{eq:phase_diff}. Let $\Omega$ be an ordering of the phases of the entries of $\mathbf{z}^{(1)}$ from the smallest to the largest, then the remaining $N-1$ points can be calculated through the recursive formula
\begin{equation} \label{eq:recursive}
z_{n}^{(m+1)}=
\begin{cases}
z_{n}^{(m)}e^{i\delta} & \text{if }n=\Omega(m) \\
z_{n}^{(m)} & \text{rest}
\end{cases}
\end{equation}
for $n=1,\ldots,N$ and $m=1,\ldots,N-1$.
By exploiting \eqref{eq:recursive}, the figure of merit $S\triangleq \sum_{n=1}^{N}z_{n}$ of such points can also be computed recursively as in lines~\ref{alg:optimal_phases:for}--\ref{alg:optimal_phases:end_for} and the optimum solution identified. Lastly, $\boldsymbol{\omega}^\star$ is computed in line~\ref{alg:optimal_phases:output} by reversing the variable change.

In terms of computational complexity, the sorting in line~\ref{alg:optimal_phases:sort} runs in $\mathcal{O}(N\log N)$. Lines \ref{alg:optimal_phases:init}, \ref{alg:optimal_phases:init_sum} and \ref{alg:optimal_phases:output} have complexity $\mathcal{O}(N)$. The loop performs $N$ scalar operations and so it has complexity $\mathcal{O}(N)$. Overall, Algorithm~\ref{alg:optimal_phases} runs in $\mathcal{O}(N\log N)$. An interesting observation is that it does not depend on the number of phase shifts $Q$ or the resolution $\delta$. 
Compare $\mathcal{O}(N\log N)$ to computing the matrix-vector product explicitly as in \eqref{eq:OMP:max_corr} with complexity $\mathcal{O}(N M)$, which in the case of the complete dictionary would result in $\mathcal{O}(N Q^{N})$. 
In the next section we introduce a faster implementation than naive OMP and combine it with Algorithm~\ref{alg:optimal_phases}.

\subsection{OMP by the Cholesky Decomposition}

The previous section provided an efficient method for computing the critical step \eqref{eq:OMP:max_corr} in OMP. For the remaining steps of OMP described in Section~\ref{sec:naive}, more efficient implementations exist  in the literature \cite{sturm2012comparison}.
We choose a common variation of OMP which exploits the Cholesky decomposition, but because step \eqref{eq:OMP:max_corr} is computed differently in our algorithm, some small changes have been made to the original algorithm. In particular, our implementation does not need to keep track of the scalar products between the columns of the dictionary and the current residue. For this reason and in order to analyze the computational complexity, we have detailed its steps in Algorithm~\ref{alg:OMP_cholesky}.
\begin{algorithm} \caption{OMP by Cholesky Decomposition} \label{alg:OMP_cholesky}
	\begin{algorithmic}[1]
		
		\Statex \textbf{inputs:} $\mathbf{y}^\text{dig}$, $L$ and $\delta$
		
		\Statex \textbf{outputs:} $\mathbf{F}^\text{RF}$ and $\mathbf{f}^\text{BB}$
		
		\State	initialize $\mathbf{r}_0=\mathbf{y}^\text{dig}$
		
		\For{$l=1,\ldots,L$}
		
		\State run Algorithm~\ref{alg:optimal_phases}, inputs: $\mathbf{r}_{l-1}$ and $\delta$, output: $\boldsymbol{\omega}^\star$ \label{alg:OMP_cholesky:phases}
		
		\If{$l=1$}
		
		\State $\mathbf{T}_l=\sqrt{N}$			
		
		\Else
		
		\State find $\mathbf{v}$ in triangular system $\mathbf{T}_{l-1} \mathbf{v} = \left[\mathbf{F}^\text{RF}\right]^\mathrm{H} \boldsymbol{\omega}^\star$ \label{alg:OMP_cholesky:v_vector}
		
		\State compose $\mathbf{T}_l = \begin{bmatrix} \mathbf{T}_{l-1} & \mathbf{0} \\ \mathbf{v}^\mathrm{H} & \sqrt{N-\|\mathbf{v}\|^2} \end{bmatrix}$ \label{alg:OMP_cholesky:cholesky}
		
		\EndIf
		
		\State compose $\mathbf{F}^\text{RF} \leftarrow [\mathbf{F}^\text{RF} \, \boldsymbol{\omega}^\star]$
		
		\State solve $\mathbf{z}$ and $\mathbf{f}^\text{BB}$ in the double triangular system $\left\{\begin{aligned}
		&\mathbf{T}_l \mathbf{z} = \left[\mathbf{F}^\text{RF}\right]^\mathrm{H} \mathbf{y}^\text{dig} \\
		&\mathbf{T}_l^\mathrm{H} \mathbf{f}^\text{BB} = \mathbf{z}
		\end{aligned}\right.$ by back-substitution \label{alg:OMP_cholesky:system}
		
		\State compute $\mathbf{r}_l = \mathbf{y}^\text{dig} -\mathbf{F}^\text{RF} \mathbf{f}^\text{BB}$ \label{alg:OMP_cholesky:residue}
		
		\EndFor
		
	\end{algorithmic}
\end{algorithm}
For more details on the connection with naive OMP check \cite{sturm2012comparison} and the references therein.

The complexity of solving the triangular systems in line~\ref{alg:OMP_cholesky:v_vector} and line~\ref{alg:OMP_cholesky:system} by back substitution is proportional to the size of the matrix, i.e., $\mathcal{O}(l^2)$. The matrix products involving the $N\times L$ matrix $\mathbf{F}^\text{RF}$, in lines~\ref{alg:OMP_cholesky:v_vector}, \ref{alg:OMP_cholesky:system} and \ref{alg:OMP_cholesky:residue}, require $\mathcal{O}(NL)$ operations. Thus, the overall complexity of running iteration $l$ in Algorithm~\ref{alg:OMP_cholesky}, when taking into account the complexity of Algorithm~\ref{alg:optimal_phases} in line~\ref{alg:OMP_cholesky:phases}, is $\mathcal{O}(N\log N+l^2+Nl)$. Using the fact that $l\leq L\leq N$, the asymptotic complexity simplifies to $\mathcal{O}(N\log N+Nl)$.
\begin{table}
	\centering
	\caption{Computational complexity in running one iteration of OMP with different implementations.}
	\begin{tabular}{lcc}
		\toprule
		& \textbf{Naive OMP} & \textbf{Cholesky OMP}  \\
		\midrule
		\textbf{Steering dictionary} & $\mathcal{O}(NQ +Nl^2)$ & $\mathcal{O}(NQ+Nl)$ \cite{sturm2012comparison} \\
		\parbox[center][2.5em][c]{.3\columnwidth}{\textbf{Complete dict.\\without Algorithm~\ref{alg:optimal_phases}}}
		& \cellcolor{red!25} $\mathcal{O}\left(NQ^{N} +Nl^2\right)$ & \cellcolor{red!25} $\mathcal{O}\left(NQ^{N}+Nl\right)$ \\
		\parbox[center][2.5em][c]{.3\columnwidth}{\textbf{Complete dict.\\with Algorithm~\ref{alg:optimal_phases}}} & $\mathcal{O}(N \log N +Nl^2)$ & \cellcolor{green!25} $\mathcal{O}(N\log N+Nl)$
	\end{tabular}
	\label{tab:complexity}%
\end{table}
Table~\ref{tab:complexity} summarizes the complexity of the discussed algorithms, and marks in green the selected variation, and in red the ones with exponential computational complexity.

\subsection{Distortion in SbS Precoding}

In general, the approximation \eqref{eq:approximation} will rarely be satisfied with strict equality, meaning SbS precoding will fail to perfectly approximate the signal transmitted by a digital array (see Section~\ref{sec:SbS_LOS} for numerical evidence).
Let $\mathbf{a}(\phi,\theta)$ be the array response, then the emitted field towards azimuth-elevation $(\phi,\theta)$ is
\begin{equation} \label{eq:field}
	z_t(\phi,\theta) \triangleq  \mathbf{a}^\mathrm{H}(\phi,\theta) \mathbf{y}_t,
\end{equation}
where $(\mathbf{y}_t,{z}_t) \in \{(\mathbf{y}_t^\text{hyb},{z}_t^\text{hyb}),(\mathbf{y}_t^\text{dig},{z}_t^\text{dig})\}$. Let there be $K$ users towards $\{(\phi,\theta)\}_{k=1}^K$. Stacking the emitted signals towards all users, we obtain the vector $\mathbf{z}_t$ of emitted signals in the LOS directions:
\begin{align}
	\mathbf{z}_t &= \mathbf{A}^\mathrm{H} \mathbf{y}_t \\
	\mathbf{A} &\triangleq
	\begin{bmatrix}
		\mathbf{a}^\mathrm{H}(\phi_1,\theta_1) & \cdots & \mathbf{a}^\mathrm{H}(\phi_K,\theta_K)
	\end{bmatrix}.
\end{align}

The emitted signals can be decomposed into a factor that is linear with respect to the data streams $\mathbf{s}_k^\mathrm{T} = \begin{bmatrix}s_{1,t} & \cdots & s_{K,t}\end{bmatrix}$, and a non-linear factor $\mathbf{d}_t$, that we call distortion:
\begin{equation} \label{eq:decomposition}
	\mathbf{z}_t = \mathbf{G} \mathbf{s}_t +\mathbf{d}_t,
\end{equation}
where $\mathbf{G}$ is the matrix of gains. 
The decomposition of \eqref{eq:decomposition} is ambiguous because for fixed $\mathbf{G}$, we can always find an appropriate distortion $\mathbf{d}_t$ that explains $\mathbf{z}_t$. For instance, we could define all signals as distortion, i.e., $\mathbf{G}=\mathbf{0}$ and $\mathbf{z}_t = \mathbf{d}_t$, but such decomposition would be uninformative. We are interested in a decomposition where most of the energy is explained by the linear term, which results in the following least square fit:
\begin{equation}
\hat{\mathbf{G}} = \argmin{\mathbf{G}}  \sum_{t=1}^T\left\|\mathbf{z}_{t}-\mathbf{G} \mathbf{s}_t\right\|^2.
\end{equation}
Assuming the number of transmissions $T$ is equal or larger than the number of streams $K$, then the solution is the well known closed-form formula $\hat{\mathbf{G}} = (\sum_{t}\mathbf{z}_t\mathbf{s}_t^\mathrm{H}) (\sum_{t}\mathbf{s}_t\mathbf{s}_t^\mathrm{H})^{-1}$,
and the distortion is computed as the remainder
\begin{equation}
	\hat{\mathbf{d}}_t = \mathbf{z}_t -\mathbf{G} \mathbf{s}_t.
\end{equation}
Trivially, for the case of a digital array \eqref{eq:signal_digital}, $\mathbf{G} = \mathbf{A}^\mathrm{H}\mathbf{F}$ and $\mathbf{d}_t = \mathbf{0}$. Unfortunately, in general SbS precoding will introduce distortion due to its non-linear nature.

\section{Numerical Results} \label{sec:numerical_results}

In the following experiments, the base station (BS) equips a uniform linear array (ULA) with 16 antenna elements and half-wavelength inter-antenna spacing. The ULA is at the origin of the coordinate system and aligned with the $y$ axis.
$K$ independent streams precoded with $K$ different beams are transmitted towards $K$ different users positioned at $K$ different azimuth angles and elevation 0. The streams are coded with a QPSK constellation with unit magnitude and the phases of the symbols are drawn from a uniform random distribution.
The goal of the following experiments is to compare the performance of three different schemes:
\begin{enumerate}
	\item Hybrid array with standard precoding. The switches remain fixed for the entire duration of the streams, and therefore the number of users that the BS can serve is bounded by the number of RFCs.
	\item Hybrid array with SbS precoding.
	\item Digital array.
\end{enumerate}

The BS's array response for an inbound ray with azimuth-elevation pair $(\phi,\theta)$ is defined in \cite{kelley1993array} as
\begin{equation}
	\mathbf{a}(\phi,\theta) =
	g(\phi,\theta)
	\begin{bmatrix}
		1 & e^{j\pi\sin\phi} & \cdots & e^{j\pi(N-1)\sin\phi_k}
	\end{bmatrix}^\mathrm{T},
\end{equation}
where $g(\phi,\theta)$ is the antenna radiation pattern of the patch antenna proposed by 3GPP \cite{3GPP2017study}. Because of the patch antennas' radiation pattern, the ULA only serves users positioned within the azimuth range $[\SI{30}{\degree},\SI{150}{\degree}]$. For the digital precoders of user $k$ \eqref{eq:signal_digital}, we choose a beam of the form
\begin{equation} \label{eq:beam}
	\mathbf{f}_k^\text{dig} = \frac{c}{g(\phi_k,0)}
	\begin{bmatrix}
		1 & e^{j\pi\sin\phi_k} & \cdots & e^{j\pi(N-1)\sin\phi_k}
	\end{bmatrix},
\end{equation}
where the factor $1/g(\phi_k,0)$ compensates for the antennas' radiation pattern, and constant $c$ is used to normalize the total radiated power to the radiated power of
an isotropic antenna with unit gain, i.e., $c = (4\pi T/\sum_{t=1}^T I_t)^{1/2}$ where $I_t=\int_{0}^{2\pi} \int_{-\frac{\pi}{2}}^{\frac{\pi}{2}}\mathbf{a}^\mathrm{H}(\phi,\theta)\mathbf{y}_t \cos \theta \,\mathrm{d}\theta \,\mathrm{d}\phi$ \cite[Eq.~(33)]{ivrlac2010toward}. 
Depending on the metric of choice, better precoding strategies for digital arrays exist in the literature \cite{lee2007high}.
These digital precoders aim to deliver similar symbol rates to all users and can also be implemented, with minor modifications, with a hybrid array with standard precoding. Precisely, for standard hybrid precoding, the number of streams/precoders is capped to the number of RFCs ($K\leq L$):
\begin{equation}
	\mathbf{F}^{\text{RF}} = 
	\begin{bmatrix}
		\tilde{\mathbf{f}}_1 & \cdots & \tilde{\mathbf{f}}_{\min(L,K)}
	\end{bmatrix}
\end{equation}
where $\tilde{\mathbf{f}}_k$ is the closest approximation of $\mathbf{f}_k^\text{dig}$ such that its entries lie in the discrete set of phase shifts \eqref{eq:set_of_phases}. Lastly, the SbS precoder is obtained by solving \eqref{opt:hybrid_precoding:LS}.

\subsection{Beampatterns}

\begin{figure}
	\includegraphics{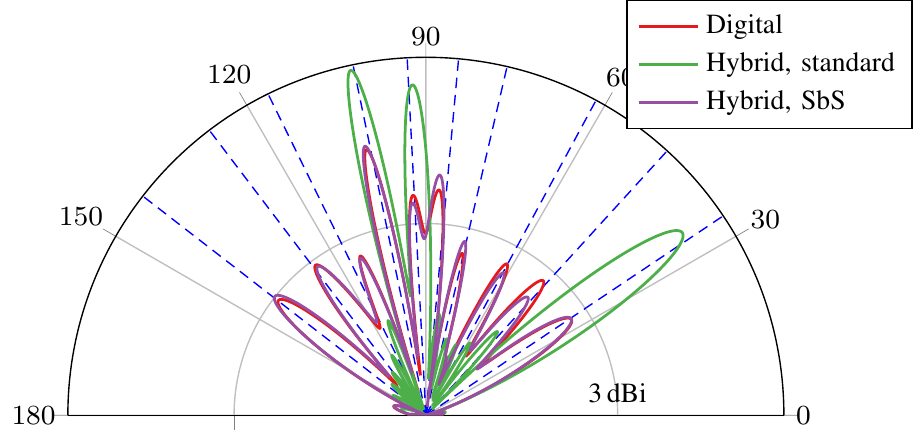}
	\label{fig:beampatterns}
	\caption{Beampatterns produced by three different types of arrays: a digital array serving 10 users, a hybrid array serving 10 users with SbS precoding, and a hybrid array using standard precoding to serve 3 users. The uniform linear array (ULA) has 16 antenna elements. Only half plane is plotted because the energy backscattered by the array is negligible. The dashed lines indicate the directions of the users.}
\end{figure}

Beampatterns are obtained by plotting $|z_t(\phi,\theta)|$ in \eqref{eq:field} as a function of azimuth, and they are useful for visualizing energy transmitted in different directions. In Fig.~\ref{fig:beampatterns}, we plot the beampatterns for the digital, hybrid and SbS precoders, when transmitting the QPSK symbols $-1$, $j$, $-j$, $-1$, $-1$, $1$, $-j$, $-1$, $-1$ and $j$, towards 10 users at azimuth angles \SI{34}{\degree}, \SI{48}{\degree}, \SI{62}{\degree}, \SI{77}{\degree}, \SI{85}{\degree}, \SI{93}{\degree}, \SI{102}{\degree}, \SI{116}{\degree}, \SI{127}{\degree} and \SI{142}{\degree}, respectively. The hybrid array employs 3 RFCs and phase shifters with \SI{45}{\degree} resolution ($Q=8$).
As expected, the beampattern of the digital precoder puts ten beams in the direction of the users. The beampattern of the SbS precoder closely matches that of the digital precoder. The hybrid array with standard precoder can only emit 3 beams because there are only 3 RFCs. Because the standard hybrid precoder serves less user, the array gain towards each user is also higher.

\subsection{Distortion Analysis} \label{sec:SbS_LOS}

The distortion introduced by SbS precoding could limit its performance. The simulated precoders \eqref{eq:beam} maximize the gain towards each user, but they also introduce 
secondary lobes in the directions of other users, causing inter-beam (or inter-user) interference.
For the distortion of SbS precoding to be negligible in comparison to hybrid or digital beamforming, it should be an order of magnitude below other types of interferences.
We derive expressions for the signal gain and interference power from the formula in \eqref{eq:decomposition}. Precisely, let $g_{k,k'}$ be the $(k,k')$-th entry of matrix $\hat{\mathbf{G}}$, then the gain for the $k$-th user is $g_{k,k}$ and the inter-beam interference from user $k'$ to user $k$ is $g_{k,k'}$, and the signal-to-interference-plus-distortion ratio ($\mathrm{SIDR}$) averaged for all users is defined as
\begin{equation}
	\mathrm{SIDR} = \frac{1}{K} \sum_{k=1}^{K}\frac{g_{k,k}^2}{P_k^\text{i} +P_k^\text{d}}
\end{equation}
where 
\begin{align}
	P_k^\text{i} &= \sum_{\substack{k'=1\\k\neq k'}}^{K} g_{k,k'}^2 &
	P_k^\text{d} &= \frac{1}{T}\sum_{t=1}^{T}|d_{k,t}|^2.
\end{align}

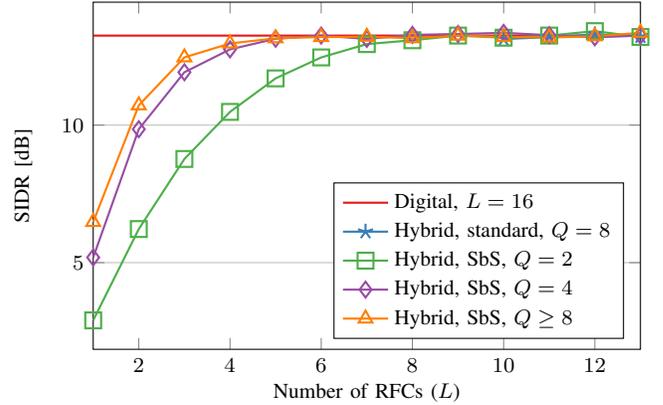
\begin{figure}
	\centering
	\begin{tikzpicture}
	\begin{axis}[
		xmin=1,xmax=13,
		ylabel={SIDR [\si{\decibel}]},
		xlabel={Number of RFCs ($L$)},
		cycle list name=myCycleList,
		legend entries = {Digital, $L=16$\\Hybrid, standard, $Q=8$\\Hybrid, SbS, $Q=2$\\Hybrid, SbS, $Q=4$\\Hybrid, SbS, $Q\geq8$\\},
		legend style={legend pos=south east,cells={align=left}},
		legend cell align=left,
	]
	\addplot+[mark=none] table[
		x=RFC,
		y=digital,
	] {./SIDRvsRFC.dat};
	\addplot table[
		x=RFC,
		y=hybrid,
	] {./SIDRvsRFC.dat};
	\addplot table[
		x=RFC,
		y=SbS_2,
	] {./SIDRvsRFC.dat};
	\addplot table[
		x=RFC,
		y=SbS_4,
	] {./SIDRvsRFC.dat};
	\addplot table[
		x=RFC,
		y=SbS_8,
	] {./SIDRvsRFC.dat};
	\end{axis}
	\end{tikzpicture}
	\caption{SIDR averaged for all users and over 100 channel realizations for three different types of arrays. The digital array and the hybrid array with SbS precoding serve 10 users simultaneously, while the hybrid array with standard precoding serves $L$ users only.  The SIDR for the hybrid array with standard precoding is only plotted for $L\geq 10$ because it cannot serve the 10 users otherwise, and its value matches that of the other schemes ($\sim\SI{13.3}{\decibel}$).}
	\label{fig:SIDR}
\end{figure}

Fig.~\ref{fig:SIDR} plots the SIDR as a function of the number of RFCs in the hybrid array. In this Monte Carlo experiment the users and the QPSK signals are drawn independently between runs. The users are distributed randomly across azimuth but always checking a minimum azimuth distance is preserved between users such that the precoder's mainlobes do not overlap.
By definition, the number of RFCs of the digital array remains constant and equal to the number of antennas ($L=16$).
As the number of RFCs increases, the SIDR of SbS precoding converges to that of the digital array. Similarly, increasing the resolution of the phase shifters also improves the SIDR of SbS precoding up to $Q=8$ (\SI{45}{\degree} resolution). We observed that for higher resolutions of the phase shifters the improvement in SIDR is negligible. 
The SIDR of the hybrid array with standard precoding is only plotted for $L\geq 10$ because it requires more RFCs than users, and its value is similar ($\sim\SI{13.3}{\decibel}$) to the other precoding schemes.

\subsection{Sum-Rate Analysis}

Thanks to serving more users, we would expect that SbS hybrid precoding achieves higher sum-rates than standard hybrid precoding. To compute the maximum achievable sum-rate, we transmit independently drawn Gaussian symbols with zero-mean and unit-variance for all users, and assume no cooperation between users.
The channel between the BS and each user is modeled as pure LOS. This is a pessimistic channel from the perspective of SbS precoding because, in general, pure LOS channels will not suffer inter-user interference due to multipath, making the distortion more dominant.
We also make the simplifying assumption that the distortion is Gaussian. Thus, for user $k$, the maximum achievable rate is given by Shannon's capacity formula
\begin{equation}
	C_k = R_k \log_2 \left(1+\mathrm{SINDR}_k\right)
\end{equation}
where $R_k$ is the symbol rate of the $k$-th stream and the $\mathrm{SINDR}_k$ (signal-to-interference-noise-and-distortion ratio) is defined as
\begin{equation}
	\mathrm{SINDR}_k = \frac{h_k g_{k,k}^2}{\sigma_\text{n}^2 +h_k P_k^\text{i} +h_k P_k^\text{d}}.
\end{equation}
Here, $h_k$ is the LOS channel gain, $\sigma_\text{n}^2$ is the noise variance and assumed equal for all users, and the rest of parameters were defined in Section~\ref{sec:SbS_LOS}. The maximum achievable sum-rate is then defined as
\begin{equation}
	\mathrm{SR} = \sum_{k=1}^{K} C_k.
\end{equation}
A Monte Carlo simulation is performed to compute the average sum-rate. 
As in the previous section, for each Monte Carlo run, the users are position randomly on a sector of \SI{120}{\degree} width in azimuth, and $h_k$ is modeled as an exponential random variable with unit mean. The number of discrete phase shifts is set to $Q=8$ and the number of RFCs is $L=4$. The noise variance at the receiver is set such that if the BS consisted of a single isotropic antenna and the channel gain was 1, then the receive SNR would be \SI{0}{\decibel}.

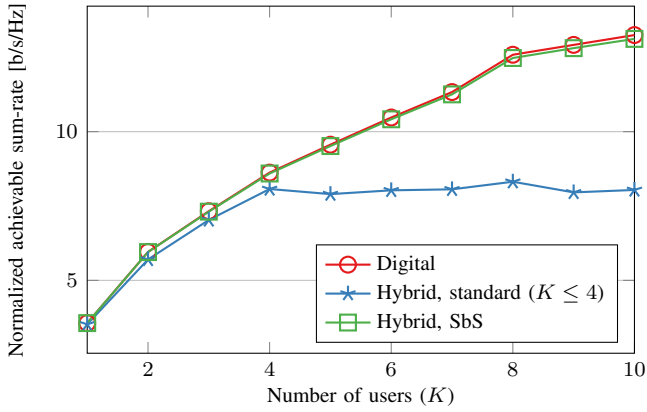
\begin{figure}
	\centering
	\begin{tikzpicture}
	\begin{axis}[
	xmin=1, xmax=10,
	xlabel={Number of users ($K$)},
	ylabel={Normalized achievable sum-rate [\si{\bit/\second/\hertz}]},
	cycle list name=myCycleList,
	legend entries = {Digital\\Hybrid, standard ($K\leq4$)\\Hybrid, SbS\\},
	legend style={legend pos=south east,cells={align=left}},
	legend cell align=left,
	]
	\addplot table[
	x=Nusers,
	y=digital,
	] {./SRvsUSERS.dat};
	\addplot table[
	x=Nusers,
	y=hybrid,
	] {./SRvsUSERS.dat};
	\addplot table[
	x=Nusers,
	y=SbS,
	] {./SRvsUSERS.dat};
	\end{axis}
	\end{tikzpicture}
	\caption{Normalized achievable sum-rate averaged over 100 channel realizations for three different types of arrays. By multiplexing independently generate streams of data to different users, the BS is able to increase its total throughput.}
	\label{fig:sum-rate}
\end{figure}
Fig.~\ref{fig:sum-rate} plots the achievable sum-rate normalized by the signal bandwidth, averaged among 100 channel realizations.
For the hybrid array with standard precoding, we can observe that its sum-rate saturates for $L\geq 4$ because the number of users served is limited by the number of RFCs. On the other hand the sum-rates of digital and SbS precoding monotonically increase with the number of users thanks to the spatial multiplexing of the streams.

\section{Conclusions}

Arrays with hybrid analog-digital architectures are an attractive solution for mmWave communications. 
However, an inherent limitation of base stations with hybrid arrays is that the number of users served in the same time-frequency slot is limited to the number of RF chains. To bypass this limitation, we proposed a precoding strategy for the hybrid array architecture of \cite{yu2018hardware}, that changes the states of the switches on a symbol-by-symbol (SbS) basis.
Thanks to SbS precoding large numbers of users can be served even when exceeding the number of RF chains. The proposed technique jointly optimizes the states of the switches and the symbols passed to the RF chains using a variation of orthogonal matching pursuit (OMP) optimized for our particular problem. Numerical experiments revealed that only 4 RF chains are necessary for making the distortion introduced by SbS precoding negligible. 
Nonetheless, further research is needed for mitigating the spectral regrowth and inter-symbol interference caused by the fast switching.

\bibliographystyle{IEEEtran}
\bibliography{IEEEabrv,./references}

\end{document}